\documentclass[11pt]{article}
\usepackage{amsfonts}
\usepackage{amsmath}
\usepackage[dvips]{graphicx}
\usepackage{amscd}
\usepackage{amssymb}
\usepackage{wrapfig}
\newtheorem{theorem}{Theorem}

\newtheorem{remark}{Remark}

\begin{document}

\title{\textbf{Zeno--line, Binodal, $T$--$\rho$ Diagram and Clusters as a new
Bose-Condensate Bases on New Global Distributions in Number Theory}}

\author{V.P. Maslov\thanks{Department of Physics, Moscow State University, Vorob'evy
Gory, 119234 Moscow, Russia. email: v.p.maslov@mail.ru} }
\date{}
\maketitle

\begin{abstract}
We present the correspondence principle between the $T$--$\rho$ diagram, the
Zeno line, and the binodal for a given interaction potential of Lennard-Jones
type. We use this correspondence further to construct a distribution of the
Bose--Einstein type for a classical gas with the help of the new notion of
Bose condensate, making it possible to decrease fractal dimension while
simultaneously preserving the number of particles. In so doing, we use new
global distributions in number theory.

\bigskip



\end{abstract}

\bigskip

\bigskip

\textit{This paper is dedicated to Stephen Smale on the occasion of his
eightieth birthday.}

\section{Introduction}

At the origin of the mathematical theory of distributions (generalized
functions) which result from the fundamental works of L. Schwartz, and also of
Gel'fand and Shilov, the so-called ``Schr\"{o}dinger problem'',
or the ``Schr\"{o}dinger conjecture'',
arose in mathematics. In the famous memoir in
which Schr\"{o}dinger introduced his equation, he gave definitions of
eigenfunctions for the discrete and the continuous spectrum. He defined the
eigenfunctions of the discrete spectrum correctly, whereas the definition of
functions corresponding to the continuous spectrum contained an inessential
error. Namely, Schr\"{o}dinger assumed that the functions corresponding to the
points of continuous spectrum are bounded at infinity. However, in fact, these
functions can have a growth which cannot exceed that of some power of $r$,
where $r$ stands for the radius, as $r\rightarrow\infty$.

I published the related counterexample in 1968 \cite{Maslov_1968}. The
mathematical proofs of the above conjecture (such proofs were presented by
Maurin, Kostyuchenko, Gel'fand, and Shilov) contained errors, which the
authors had found themselves when studying my counterexample. A counterexample
in the case of an absolutely continuous spectrum was given in 1993 by the
author of the present paper and S.~Molchanov and was reported in a plenary
talk at the Congress on Mathematical Physics. The so-called Steklov problem
for polynomials was solved simultaneously (in the negative;
see~\cite{Maslov_Molchanov_1993}).

Certainly, I had no doubt that Schr\"odinger's definition of eigenfunctions is
inexact; however, I presented the counterexample only after the problem in
question passed to the area of mathematics. The point is that there are many
well-known paradoxes in physics. On the contrary, there must be no paradoxes
in mathematics (in mathematical physics). The well-known Gibbs paradox was
regarded as a physical paradox, and fifteen Nobel prize laureates tried to
solve it. However, two famous mathematicians, von Neumann and Poincar\'e, also
tried to solve this paradox. After the exhausting paper by V.~V.~Kozlov
\cite{Kozlov_2002}, the Gibbs paradox finally moved to the mathematical area.
Kozlov proved the following fact~\cite{Kozlov_2009}: it follows from the
Poincar\'e model concerning the dynamics of collisionless gas in a rectangular
parallelepiped with mirror walls that, if an interior wall between two parts
of the vessel disappears, then the entropy increases stepwise.

However, it is clear that, if we evaluate the density, i.e., the number of
particles in a unit volume, then the entropy, both as the logarithm of the
number of possible variants and as the Kolmogorov complexity, is preserved.

The problem of correctly defining the notion of ideal gas became now mature in
mathematics as well.

It should be noted that the microcanonical Gibbs distribution holds (see Theorem~2).

A completely different approach not connected with the microcanonical Gibbs
distribution was proposed simultaneously by Green, Kirkwood, and others. This
approach was studied in great detail by N.~N.~Bogolyubov. However, in his
construction, Bogolyubov used the conjecture of chaos preservation stated by
Mark Kac. A counterexample to this conjecture was constructed by the author of
the present paper and O.~Shvedov in \cite{Masl_Shvedov}.

\section{Partition of integers and revision of the Bose--Einstein
distribution}

The relationship between the Bose--Einstein distribution and number theory was
studied, in particular, in \cite{Freiman_Vershik}, \cite{Masl_Naz_83_2},
\cite{Masl_Naz_83_3}, \cite{Masl_Naz_83_6}, \cite{Vershik_1},\cite{Vershik_2}.
Let us consider two examples.

As an example of a simple model of Bose condensate, consider Koroviev's trick
(well known from M.~Bulgakov's novel ``Master and Margarita'')
of scattering money bills in a variety theater's
audience. According to number theory and Kolmogorov complexity, if Koroviev
had one million bills and the number of spectators were ten thousand, then
only one thousand of them would get bills (see~\cite{Masl_Naz_83_3}). The
other nine thousand would not get a single bill (and presumably would die of
hunger). This is exactly a model of Bose condensate. But if the spectators
united into groups of ten and agreed to divide the bills between themselves,
then, figuratively speaking, no one would die, i.e., the number of spectators
(or particles) would be preserved. But the association of the
spectators into groups would mean a constraint on their degrees of freedom,
just as the association of particles into clusters.

This leads to two conclusions.

First, from the mathematical point of view, this example is equivalent to the
existence of a two-dimensional Bose condensate, but this refutes the
physicists' postulate (dating back to Einstein) that no two-dimensional
condensate does exists. It turns out that, in the Bose--Einstein distribution,
it is necessary to add a special term taking into account the fact that the
number of particles is finite. Simultaneously, it provides the asymptotic
distribution function (previously not known) for the number of bills obtained
by groups of spectators.

Second, the Bose condensate can be regarded as an association of dimers,
trimers, and clusters, not only of ``frozen''
\ particles precipitating to the Bose condensate. And this applies to a
classical gas, not a quantum, one. Thus, under the condition $N=const$, the
Bose--Einstein distribution can be regarded as a distribution for a classical
gas and the degeneracy temperature as the critical temperature
\cite{Maslov_83_5}.\medskip

\noindent\textbf{Example 1. }By way of an example, similar to approach of
molecular dynamics, let us consider a set of unnumbered billiard balls of unit
mass and the same color. First, consider one billiard ball and launch it from
some (arbitrary) point with velocity not exceeding a certain sufficiently
large value $v$, i.e., with energy not exceeding $\ v^{2}/2$. However, since
the computer has certain accuracy, it follows that the energy of the ball will
take a finite integer number $s$ of values in the interval $[0,v^{2}/2]$:
$\lambda_{i}=iE_{0}$, $i=1,\dots,s$, of energies where $E_{0}$ corresponds to
this accuracy. Thus, we obtain the spectrum of energy values which can be
regarded as a self-adjoint diagonal matrix of order $s$, where $s\gg1$.

By assigning such a discrete set of energies to a ball, we obtain the
wave--particle correspondence in classical mechanics, because the resulting
matrix is unitarily equivalent to any operator $\widehat{L}$ with such a
spectrum in a Hilbert space~$H$.

How many balls must be launched so that the computer is not able to determine
their initial data?

The spectrum corresponding to $N$ balls can be obtained by considering the
tensor product of $N$ Hilbert spaces $H$ and the corresponding spectrum of
operator
\[
\widehat{L}_{N}=\widehat{L}\otimes\widehat{L}\otimes\dots\otimes\widehat
{L}\qquad\text{(}N\text{ times). }
\]
The eigenvalues of this operator are of the form
\[
{\mathcal{E}}=\sum_{i=1}^{s}N_{i}\lambda_{i}
\]
If we only consider the eigenfunctions, symmetric with respect to the
permutation of the particles, of this operator, which corresponds to the
identity of the balls, then the eigenvalue ${\mathcal{E}}=\sum_{i=1}^{s}
N_{i}\lambda_{i}$ is of multiplicity equal to the number of all possible
variants of the solution of the problem
\begin{equation}
{\mathcal{E}}=\sum_{i=1}^{s}N_{i}\lambda_{i},\qquad\sum_{i=1}^{s}N_{i}=N.
\label{JF_1}
\end{equation}

If $d$  is  the topological dimension, then
\begin{equation}\label{JF_1a}
\lambda_i= \frac{(i+d-1)!}{i!(d-1)!}.
\end{equation}
The binary logarithm of solutions of these Diophantine equations is called Hartley entropy. Let us consider a problem
\begin{equation}
\sum_{i=1}^{s}N_{i}\lambda_{i}\leq {\mathcal{E}}, \qquad\sum_{i=1}^{s}N_{i}=N .
\label{JF_1b}
\end{equation}
The entropy of  problems~\eqref{JF_1}  and~\eqref{JF_1b} coincide  up to $\sqrt{N\ln
\,N}$ (the accuracy up to which we solve these  problems).

This consideration allows to extrapolate the above  theory of integers to non-integral dimensions.
We shall consider  relations of the form
\begin{equation}
\sum_0^\infty\frac{\Gamma(d+i)}{\Gamma(i+1)\Gamma(d)}N_i\leq {\mathcal{E}},
\label{JF_1c}
\end{equation}
\begin{equation}
\sum_0^\infty N_i=N,
\label{JF_1d}
\end{equation}
where $\Gamma(d)$ is  the $\Gamma$-function,  and the number of solutions satisfying inequality~\eqref{JF_1c} and equality~\eqref{JF_1d} for non-integer $d$ (``fractial dimension'').

Since the initial set of energies is ``without
preferences'', i.e., is in a general position, then all the
multiplicities corresponding to (\ref{JF_1}) are equiprobable. The computer
calculation time is related to the computer accuracy $E_{0}$ with respect to
energy. The problem is how to determine the number $N$ for which, at a
particular instant of time, the computer cannot recover the initial data in
view of inaccuracy of the classical pattern or that of the quantum mechanical
pattern (which is more accurate, but more cumbersome); the latter pattern, in
turn, is not accurately described by an interaction of Lennard-Jones type.

Thus, we can draw the following conclusion. The initial data in classical and
quantum mechanical problems are discarded due to external noise. As a result,
the problem is reduced to the distribution of $\{N_{i}\}$ in (\ref{JF_1}). In
this problem, we assume \textit{a priori} that the initial data are discarded
and, therefore, so is the numbering of classical particles. Although the
textbook \cite{Landau_Kv_Mech} falsely interpreted the difference between the
quantum mechanical pattern and the classical one\footnote{The authors explain
the identity principle for particles as follows: ``In
classical mechanics, identical particles (such as electrons) do not lose their
'identity' despite the identity of their physical properties. ... we can
'number' them and then observe the motion of each of them along its
trajectory; hence, at any instant of time, the particles can be identified ...
In quantum mechanics, it is not possible, in principle, to observe each of the
identical particles and thus distinguish them. We can say that, in quantum
mechanics, identical particles completely lose their 'identity'
\cite{Landau_Kv_Mech}, p.~252. (But, as a matter of fact, if
the initial data for the Cauchy problem does not possess a symmetry property,
then the situation in quantum mechanics does not differ from that in classical
mechanics).}, it, nevertheless, gave a valid interpretation of the numbering
of identical balls. Therefore, we can take symmetric eigenfunctions for
$\widehat{L}_{N}$.

Therefore, it only remains to obtain the distribution of number $N_{j}$ of
particles using relations~(\ref{JF_1}). If $s\gg N$, then relation~(\ref{JF_1})
can be expressed as
\begin{equation}
E_{0}\sum_{i=1}^{\infty}iN_{i}={\mathcal{E}},\qquad\sum_{i=1}^{\infty}N_{i}=N.
\label{JF_2}
\end{equation}
These relations coincide with those in the classical number-theoretic problem
under the condition that ${\mathcal{E}}/E_{0}$ is an integer, which, of
course, is of no importance in the asymptotics in $s\rightarrow\infty$ and
$N\rightarrow\infty$.

Thus, since the noise component has prevented us to recover the initial
data\footnote{As to the well-known discussion between Boltzmann and some
mathematics~\cite{Boltzman}, of course, if the particles of the gas are
distinguishable and can be numbered, then they can also be turned back and
returned to their initial state.} and the number of particles is preserved and
so is the total energy~$\mathcal{E}$ (at least, the latter is not increased),
without giving any preference we assume all the variants satisfying the
relation
\begin{equation}
\sum_{i=1}^{\infty}iN_{i}\leq\frac{\mathcal{E}}{E_{0}}\,,\qquad
\sum_{i=1}^{\infty}N_{i}=N \label{JF_3}
\end{equation}
to be equiprobable.

Under this approach, we can also take into account collision of billiard
balls. Indeed, the initial energy of all balls can only decrease due to
friction and the passage of kinetic energy into thermal energy during
collisions. The number of balls will remain the same and the total energy will
not exceed the initial energy~${\mathcal{E}}$.

There is a similar problem in number theory. Let $n$ be a positive integer. By
a partition of $n$ we mean a way to represent a natural number $n$ as a sum of
natural numbers. Let $p(n)$ be the total number of partitions of $n$, where
the order of the summands is not taken into account, i.e., partitions that
differ only in the order of summands are assumed to be the same. The number
$p_{k}(n)$ of partitions of a positive integer $n$ into $k$ positive integer
summands is one of the fundamental objects of investigation in number theory
(see \cite{Masl_Naz_83_2}, \cite{Vershik_1}, \cite{Vershik_2}).

For a given partition, denote a number of summands (in the sum) equal to $1$
by $N_{1}$, the number of summands equal to 2 by $N_{2}$, etc., and the number
of summands equal to $i$ by $N_{i}$. Then $\sum N_{i}=k$ is the number of
summands, and the sum $\sum iN_{i}$ is obviously equal to the partitioned
positive integer. Thus, we have
\begin{equation}
\sum_{i=1}^{\infty}iN_{i}=n,\qquad\sum_{i=1}^{\infty}N_{i}=k, \label{NT_1}
\end{equation}
where $N_{i}$ are natural numbers not exceeding $k$.

These formulas can readily be verified for the above example. Here, all
families $\{N_{i}\}$ are equiprobable.

A new global distribution for $N_{i}\leq k$ and $\sum N_{i}=k$ is determined
from the relations
\begin{equation}
\sum_{i=1}^{n}\left(  \frac{1}{e^{b(i+\kappa)}-1}-\frac{k}{e^{bk(i+\kappa)}
-1}\right)  =k,\qquad\sum_{i=1}^{n}\left(  \frac{i}{e^{b(i+\kappa)}-1}
-\frac{ik}{e^{bk(i+\kappa)}-1}\right)  =n, \label{NT_2}
\end{equation}
where $b>0$ and $\kappa>0$ are constants defined from (\ref{NT_2}), $n/k$ is
sufficiently large, numbers $n$ and $k$ are also large, and, by using the
Euler--Maclaurin summation formula, we can pass to the integrals (for the
estimates of this passage, see~\cite{Masl_Naz_84_1}):
\begin{align}
{\displaystyle\int\limits_{0}^{\infty}}
\left(  \frac{1}{e^{b(x+\kappa)}-1}-\frac{k}{e^{bk(x+\kappa)}-1}\right)  \,dx
&  \cong k,\label{NT_3}\\
{\displaystyle\int\limits_{0}^{\infty}}
\left(  \frac{x}{e^{b(x+\kappa)}-1}-\frac{kx}{e^{bk(x+\kappa)}-1}\right)
\,dx  &  \cong n. \label{NT_4}
\end{align}

It can be proved that $\kappa=0$ gives number $k_{0}$ with satisfactory
accuracy. Hence,
\begin{equation}
k_{0}=
{\displaystyle\int\limits_{0}^{\infty}}
\left(  \frac{1}{e^{bx}-1}-\frac{k_{0}}{e^{bk_{0}x}-1}\right)  \,dx.
\label{kor}
\end{equation}

Consider the integral (with the same integrand as in (\ref{kor})) taken from
$\varepsilon$ to $\infty$ and perform the change of variables $bx=\xi$ in the
first term and $bk_{0}x=\xi$ in the second term. Passing to the limit as
$\varepsilon\rightarrow0,$ we derive:
\begin{align}
k_{0}  &  =\frac{1}{b}\lim_{\varepsilon\rightarrow+0}\left(
{\displaystyle\int\limits_{\varepsilon b}^{\infty}}
\frac{\,d\xi}{e^{\xi}-1}-
{\displaystyle\int\limits_{\varepsilon bk_{0}}^{\infty}}
\frac{\,d\xi}{e^{\xi}-1}\right)  =\frac{1}{b}\lim_{\varepsilon\rightarrow+0}
{\displaystyle\int\limits_{\varepsilon b}^{\varepsilon bk_{0}}}
\frac{\,d\xi}{e^{\xi}-1}\nonumber\\
&  =\frac{1}{b}\lim_{\varepsilon\rightarrow+0}
{\displaystyle\int\limits_{\varepsilon b}^{\varepsilon bk_{0}}}
\frac{\,d\xi}{\xi}=\frac{1}{b}\lim_{\varepsilon\rightarrow+0}\left\{
\ln(\varepsilon bk_{0})-\ln(\varepsilon b)\right\}  =\frac{1}{b}\ln k_{0}.
\label{NT_5}
\end{align}

On the other hand, making the change $bx=\xi$ in (\ref{NT_4}), we have:
\begin{equation}
\frac{1}{b^{2}}\int_{0}^{\infty}\frac{\xi\,d\xi}{e^{\xi}-1}\cong n.
\label{NT_5'}
\end{equation}
Relations (\ref{NT_5}) and (\ref{NT_5'})\ imply
\begin{equation}
b=\sqrt{\frac{1}{n}\int_{0}^{\infty}\frac{\xi\,d\xi}{e^{\xi}-1}},\qquad
k_{0}=\frac{\sqrt{6n}}{2\pi}\ln n(1+o(1)). \label{NT_6}
\end{equation}

Let us now find the next term of the asymptotics by setting
\[
k_{0}=c^{-1}n^{1/2}\ln c^{-1}n^{1/2}+\alpha n^{1/2}+o(n^{1/2}),\qquad
\text{where}\quad c=\frac{2\pi}{\sqrt{6}}\,.
\]
Using the formula
\[
k_{0}=c^{-1}n^{1/2}\ln k_{0}
\]
and expanding $\ln k_{0}$ in
\[
\text{$\frac{\alpha}{c^{-1}\ln c^{-1}n^{1/2}}\,,$}
\]
we derive
\[
\text{$\alpha=-2\ln\frac{c}{2}$.}
\]
Thus, we have obtained the Erd\"{o}s formula \cite{Erdos}.

We note that, in the Koroviev's trick mentioned above, the problem is put not
quite similar to that in number theory on partitions of a positive integer $n$
into $k$ positive integer summands. Namely, the problem is put in the
following way - for a given positive integer number $\overline{k},$ find
$k\leq\overline{k}$ for which the number $p_{k}(n)$ of partition variants is
maximal. We have: (i) if $\overline{k}\leq k_{0},$ then the maximal number of
variants corresponds to $k=\overline{k}$; (ii) if $\overline{k}>k_{0}$, then
the maximal number of variants is achieved for $k=k_{0}$. This is precisely
the Bose-condensate
(see \cite{Hirsch_Smale_2004},~\cite{Maslov_Chaos_2009},~\cite{Smale_2000}).

\noindent\textbf{Example 2.} The above case corresponds to the consideration of the
topological dimension $d=2.$ Consider now the one-dimensional case of a Bose
condensate, which is important in physical problems. In the notation used in
statistical physics, $\mathcal{E}$ is the energy, $i\varepsilon$ are energy
levels, $k$ corresponds to a number $N$ of particles and $n$ corresponds to
$\mathcal{E}/{\varepsilon}$.

Define constants $b$ and $\kappa$ from the following relations:
\begin{align}
{\displaystyle\int\limits_{0}^{\infty}}
\xi\left(  \frac{1}{e^{b(\xi+\kappa)}-1}-\frac{N}{e^{bN(\xi+\kappa)}
-1}\right)  d\sqrt{\xi}  &  =\frac{\mathcal{E}}{\varepsilon}\cong n,
\label{NT8}\\
{\displaystyle\int\limits_{0}^{\infty}}
\left(  \frac{1}{e^{b(\xi+\kappa)}-1}-\frac{N}{e^{bN(\xi+\kappa)}-1}\right)
d\sqrt{\xi}  &  =N.
\end{align}
Just as in Example 1, we take $n,$ $N$ and $n/N$ to be sufficiently large,
$N_{cr}=k_{0}.$

First, since
\begin{equation}
n\cong\frac{1}{2}
{\displaystyle\int\limits_{0}^{\infty}}
\frac{\sqrt{\xi}\,d\xi}{e^{b\xi}-1}=\frac{1}{2b^{3/2}}
{\displaystyle\int\limits_{0}^{\infty}}
\frac{\,\sqrt{\xi}d\xi}{e^{\xi}-1}, \label{NT9}
\end{equation}
we have:
\begin{equation}
b=\left(  \frac{1}{2n}
{\displaystyle\int\limits_{0}^{\infty}}
\frac{\,\sqrt{\xi}d\xi}{e^{\xi}-1}\right)^{\frac{2}{3}}.
\end{equation}
For $N_{cr}$ (corresponding to $\kappa=0$), we obtain the following relation:
\begin{align}
N_{cr}  &  =
{\displaystyle\int\limits_{0}^{\infty}}
\left(  \frac{1}{e^{b\xi}-1}-\frac{N_{cr}}{e^{bN_{cr}\xi}-1}\right)
\,d\sqrt{\xi}\label{NT12'}\\
&  =\frac{1}{\sqrt{b}}
{\displaystyle\int\limits_{0}^{\infty}}
\left(  \frac{1}{e^{\xi^{2}}-1}-\frac{N_{cr}}{e^{\xi^{2}N_{cr}}-1}\right)
\,d\xi.\nonumber
\end{align}
Subtracting $1/\xi^{2}$ from both terms in (\ref{NT12'}):
\begin{align}
N_{cr}  &  =\frac{1}{\sqrt{b}}
{\displaystyle\int\limits_{0}^{\infty}}
\left(  \frac{1}{e^{\xi^{2}}-1}-\frac{1}{\xi^{2}}\right)  d\xi+\frac{1}
{\sqrt{b}}
{\displaystyle\int\limits_{0}^{\infty}}
\left(  \frac{1}{\xi^{2}}-\frac{1}{\xi^{2}(1+\frac{N_{cr}}{2}\xi^{2})}\right)
\,d\xi\nonumber\\
&  -\frac{1}{\sqrt{b}}
{\displaystyle\int\limits_{0}^{\infty}}
\left(  \frac{N_{cr}}{e^{N_{cr}\xi^{2}}-1}-\frac{N_{cr}}{N_{cr}\xi^{2}
(1+\frac{N_{cr}}{2}\xi^{2})}\right)  \,d\xi,
\end{align}
and using relations
\begin{align}
&  \frac{1}{\sqrt{b}}
{\displaystyle\int\limits_{0}^{\infty}}
\left(  \frac{N_{cr}}{e^{N_{cr}\xi^{2}}-1}-\frac{N_{cr}}{N_{cr}\xi^{2}
(1+\frac{N_{cr}}{2}\xi^{2})}\right)  \,d\xi\\
&  =\sqrt{\frac{N_{cr}}{b}}
{\displaystyle\int\limits_{0}^{\infty}}
\left(  \frac{1}{e^{\eta^{2}}-1}-\frac{1}{\eta^{2}(1+\frac{\eta^{2}}{2}
)}\right)  d\eta,\nonumber
\end{align}
\begin{equation}
\frac{1}{\eta^{2}(1+\frac{\eta^{2}}{2})}=\frac{1}{\eta^{2}}-\frac{1}
{2(1+\frac{\eta^{2}}{2})},
\end{equation}
we come to the quadratic equation
\begin{equation}
(\sqrt{N_{cr}})^{2}-W\sqrt{N_{cr}}+W=0
\end{equation}
with respect to $\sqrt{N_{cr}}$, where
\begin{equation}
W=(2n)^{1/3}\left(  \int_{0}^{\infty}\frac{\sqrt{\xi}\,d\xi}{e^{\xi}
-1}\right)  ^{-1/3}
{\displaystyle\int\limits_{0}^{\infty}}
\left(  \frac{1}{\xi^{2}}-\frac{1}{e^{\xi^{2}}-1}\right)  \,d\xi>0.
\end{equation}
Solving this equation, we obtain:
\begin{equation}
N_{cr}=\frac{W^{2}}{4}\left(  1+\sqrt{1-\frac{4}{W}}\right)  ^{2}
\end{equation}
This implies that, for large $n$,
\begin{equation}
N_{cr}\approx c^{2}n^{2/3}
\end{equation}
where
\[
c=\frac{\int_{0}^{\infty}\left(  \frac{1}{\xi^{2}}-\frac{1}{e^{\xi^{2}}
-1}\right)  \,d\xi}{\left(  \frac{1}{2}\int_{0}^{\infty}\frac{\sqrt{\xi}
\,d\xi}{e^{\xi}-1}\right)  ^{\frac{1}{3}}}
\]

We stress that $N_{cr}$ determines the transition to the Bose condensate and
plays the same role as $k_{0}$ (see (\ref{kor})) in number theory. This
problem in number theory is connected with the Waring problem.

\bigskip

\begin{center}
* \ * \ *
\end{center}

\bigskip

In statistical calculations of the number of inhabitants in a town, the
permutation between a child and an old man does not change the total number of
inhabitants. Hence, from the point of view of the statistics of the given
calculation, they are indistinguishable. From the point of view of the
experimenter who observes the molecules of a homogeneous gas using an atomic
microscope, they are indistinguishable. He counts the number of molecules
(monomers) and, for example, of dimers in a given volume. Dimers constitute
7\% in the total volume of the gas (according to J.M.Calo). This means that
the experimenter does not distinguish individual monomers just as dimers and
counts their separate numbers. His answer does not depend on the method of
numbering the molecules.

These obvious considerations are given for the benefit of those physicists who
relate the fact that quantum particles are indistinguishable with the
impossibility of knowing the world. I do not intend to argue with this
philosophical fact, but wish to dwell only on mathematics and statistics and
distributions related to the number of objects
(compare~\cite{Cucker_Smale_2004},~\cite{Masl_Lingvostat}).

The existence of the $\lambda$--point in Helium-4 is not a quantum effect.
According to the experimental data, the compressibility factor $Z$ decreases
to $Z=0.00763$, and hence the fractal dimension decreases to almost 2 (see
below). The derivative of energy with respect to $T$ contains the derivative
with respect to the chemical potential as a summand.

This term yields a logarithmic singularity as $N\to\infty$ (just as
in~\cite{Maslov_Chaos_2009}). Moreover, if the pressure is equal to
approximately $0.05$ atm., then there is no singularity in the density of the gas.

An example of a nonholonomic constraint (cluster) is a sphere containing
another sphere gliding on two parallel-oriented skates fastened to its
diameter. Such a collection of gliding spheres inscribed in each other is an
example of a nonholonomic cluster. Here the larger sphere must coincide with
all the spherical volume of a drop (see below), and the number of spheres must
correspond to the number of constraints. As temperature decreases below
$T=2.172$, there will be an increase in nonholonomic constraint and decrease
in the number of degrees of freedom according to the first parameter.

\begin{remark}\rm
In author's papers \ \cite{Masl_Quaisi_1}, \ \cite{Masl_Quaisi_2},
\cite{Masl_Quaisi_3} it is shown that superfluidity is not a pure quantum
effect, but a semiclassical one.
\end{remark}

By the Feynman scheme (as is well known, he treated a positron as an electron
moving back with respect to time), a particle which reflected when moving back
with respect to time collided at the boundary with a particle which moved
directly with respect to time. These two particles annihilated and gave a part
of energy to the boundary. One can readily imagine what does it mean after
processing some film inversely (back with respect to time). The tangential
component, under this understanding of reflection, corresponds to the flow of
the ``pair'' along the boundary. Therefore,
to take the boundary of the vessel into account (especially if we consider a
gas in which the convection phenomenon takes place\footnote{In other words,
masses of gas move.}), we must take the above considerations into account.

From the experiment it is well known that if the radius of the capillary is
increased, then the superfluidity disappears. Namely, it disappears because
collisions with the wall lead to a loss of energy. However, the narrower is
the capillary, the higher the first transversal mode occurs in a narrow
capillary. Thus, the energy criterion for which the superfluidity disappears
depends on the radius of the capillary.

Therefore, to explain superfluidity in a vessel, we must proceed with the
ultrasecond quantization, and introduce the operators of creation and
annihilation of dimers \cite{Maslov_Kvantovan_book}.

\section{Correspondence Principle between the $T$--$\rho$ Diagram and the
Interaction Potential of Lennard--Jones type}

The Van de Waals law of corresponding states establishes the correspondence
between different gases. The most accurate calculations can be performed using
the methods of molecular dynamics, which required a large amount of computer
time. Moreover, in these methods, it is necessary to know the potential of
interaction between particles.

Here we use the classical scattering problem to establish the law of
correspondence between the given interaction potential and the diagram $\rho
$--$T$, where $\rho$ is the density and $T$ is the temperature. This was shown
heuristically in \cite{Maslov_Arx_2009} and in \cite{MZ_87_5}.

The isotropy principle (of symmetry in all directions) is one of the key
principles of molecular physics. It must also be formulated rigorously in
mathematical terms as the isotropy principle in the theory of Kolmogorov
turbulence or the Born--Karman conditions in the theory of crystals (the
problem of the crystal volume finiteness problem), especially because of rapid
development of the computer molecular dynamics similar to the computer
anisotropic turbulence \cite{Ishihara},~\cite{Kaneda}.

The usual argument in molecular physics involves the symmetry of the average
motion of the molecules in all six directions. Therefore, $1/12$ of all
particles move toward one another. Since there are three axes, it follows that
$1/4$ of all molecules collide.

In the two-body scattering problem, the Lennard-Jones interaction potential
\begin{equation}
\Phi(r^{\prime},r")=4\varepsilon\left(  \frac{a^{12}}{\left\Vert r^{\prime
}-r^{\prime\prime}\right\Vert ^{12}}-\frac{a^{6}}{\left\Vert r^{\prime
}-r^{\prime\prime}\right\Vert ^{6}}\right)  , \label{JF_4}
\end{equation}
is usually considered. Here, $\varepsilon$ is an energy of a depth of the
well, $a$ is an effective radius and $\left\Vert r^{\prime}-r^{\prime\prime
}\right\Vert $ is the distance between two particles with radius vectors
$r^{\prime},r^{\prime\prime}$. In the absence of an external potential, the
two-particle problem reduces to the one-dimensional radial-symmetric one. As
it is well known \cite{Landau_Mech}, in this problem, two quantities (energy
$E$ and momentum $M)$ are conserved. In the scattering problem, it is
convenient to consider, instead of the momentum $M$, another preserved
constant, namely, the impact parameter $B$, so that $M=\sqrt{E}B$.

The scattering of two particles of equal mass may lead sometimes to the
formation of a pair. If, simultaneously, slight friction or viscosity occurs
and a small part of the energy is dissipated to radiation (of photons), to
noise (of phonons), then, in this formation, it is natural to preserve the
original preserved (up to infinitely small losses) quantities: the impact
parameter $B$ and the energy $E$.

These heuristic considerations led me to the correspondence principle
presented below.

At each point $r\in\mathbb{R}^{3},$ the dressed (or ``thermal'')
potential $\varphi(r)$ is attractive. In addition,
because the volume $V$ is a large parameter, it follows that if
\[
\varphi(r)=\Psi\left(  \frac{a\left\Vert r\right\Vert ^{2}}{V}\right)
\]
is expanded in terms of $1/V$, then
\begin{equation}
\Psi\left(  \frac{a\left\Vert r\right\Vert ^{2}}{V}\right)
=C_{1}+\frac{C_{2}\left\Vert r\right\Vert ^{2}}{V}+O\left(
V^{-\frac{4}{3}}\right)  ,\text{ \ \ \ }C_{2}>0 \label{JF_5}
\end{equation}
Since
\begin{equation}
\left\Vert r^{\prime}\right\Vert ^{2}+\left\Vert r^{\prime\prime}\right\Vert
^{2}=\frac{\left\Vert r^{\prime}-r^{\prime\prime}\right\Vert ^{2}}{2}
+\frac{\left\Vert r^{\prime}+r^{\prime\prime}\right\Vert ^{2}}{2},
\label{JF_6}
\end{equation}
we can, just as in~\cite{Maslov_TMF_2009}, separate the variables in the
two-particle problem and obtain the scattering problem for pairs of particles
and the problem of their cooperative motion for $r^{\prime}+r^{\prime\prime}$.

Then, in the two-body scattering problem, an attractive quadratic potential
(inverted parabola) is added to the Lennard-Jones interaction potential and
the Hamiltonian of the scattering problem has the form:
\begin{equation}
H=\frac{p^{2}}{4m}+W_{eff}(\left\Vert r^{\prime}-r^{\prime\prime}\right\Vert
),\text{ \ \ \ \ \ \ }p^{2}=\left\Vert p^{\prime}-p^{\prime\prime}\right\Vert
^{2}. \label{JF_7'}
\end{equation}
where $p^{\prime},$ $p^{\prime\prime}$ are momenta of two particles with
radius vectors $r^{\prime},$ $r^{\prime\prime}$ and
\begin{equation}
W_{eff}(r)=\frac{M^{2}}{r^{2}}+u(r),\text{ \ \ \ \ \ }u(r)=4\varepsilon\left(
\frac{a^{12}}{r^{12}}-\frac{a^{6}}{r^{6}}\right)  -\alpha r^{2}, \label{JF_8}
\end{equation}
is the effective potential (for short, we denote $\left\Vert r^{\prime
}-r^{\prime\prime}\right\Vert \equiv r)$. Here, $\varepsilon$ is the energy of
the well depth, $a$ is the effective radius and $\alpha=\frac{C_{2}}{V}$.

Since Hamiltonian $H$ is equal to the total energy, we have
\begin{equation}
E_{total}=\frac{p^{2}}{4m}+W_{eff}(r)=\frac{p^{2}}{4m}+\frac{B^{2}E_{total}
}{r^{2}}+u(r), \label{JF_9}
\end{equation}
where $B$ is the impact parameter. From (\ref{JF_9}) we derive:
\begin{equation}
E_{total}=\frac{p^{2}}{4m(1-B^{2}/r^{2})}+\frac{u(r)}{1-B^{2}/r^{2}}.
\label{ki'}
\end{equation}
Thus, the Hamiltonian splits into two parts: $r<B$ and $r>B$ divided by
barrier $r=B.$ One of them (for $r<B$) is attractive and the other one is
repulsive (for $r>B$). As the repulsive barrier, we can consider precisely
that noise which creates infinitely small viscosity or friction.

For $r<B,$ the first term in (\ref{ki'}) is negative while the second term is
positive whenever $a<r<B.$ The greater the velocity, the less is the energy.
The mean velocity is the temperature. However, to study the penetration
through the barrier of the incident particle, we must plot $E$ along the $y$
axis and turn the wells upside down. Then the minimum becomes the barrier and
the maximum becomes the depth of the well.

In addition to the attraction problem, there is also the reflection problem
for $r>B,$ $r\leq a$. It is separated by barrier $r=B$ and creates repulsive
particles. The repulsive particles obstruct the motion of the particles with
$r<B.$ This is an analog of the reflection that creates ``viscosity'' in a collisionless plasma.

In the two-body scattering problem considered above, let us analyse the
potential energy
\begin{equation}
E(r)=4\varepsilon\left(  \frac{a^{6}}{r^{6}}-\frac{a^{12}}{r^{12}}-\alpha
r^{2}\right)  \left(  \frac{B^{2}}{r^{2}}-1\right)  ^{-1}, \label{JF_12}
\end{equation}
in the attractive case (i.e. for $r<B).$ Here, $\varepsilon$ is the depth of
the well, $a$ is the effective radius and $B$ is the impact parameter. By
replacing
\[
\frac{r}{a}=r^{\prime},\qquad\frac{B}{a}=\widetilde{B},
\]
we get rid of $a$. In what follows, both the wave and the prime will be omitted.

For a given $B$ and the given potential (\ref{JF_12}), minimum $E_{\min}$ and
maximum $E_{\max}$ of $E(r)$ are achieved at points defined by condition
\begin{equation}
\frac{dE}{dr}=0. \label{JF_13}
\end{equation}
For each $\alpha\leq\alpha_{0},$ there exists $B_{0}(\alpha)$ such that
$E_{\max}=E_{\min}$ and, therefore,
\begin{equation}
\frac{d^{2}E}{dr^{2}}=0. \label{JF_14}
\end{equation}
On the graph $(\alpha,E)$ the relation $E_{\max}=E_{\min}$ determines the
analog of the Zeno line.

Let us now represent the Zeno line on the graph $(\rho,T)$.

\begin{figure}[ht]
\begin{center}
\includegraphics[height=3.2387in,width=3.1635in]{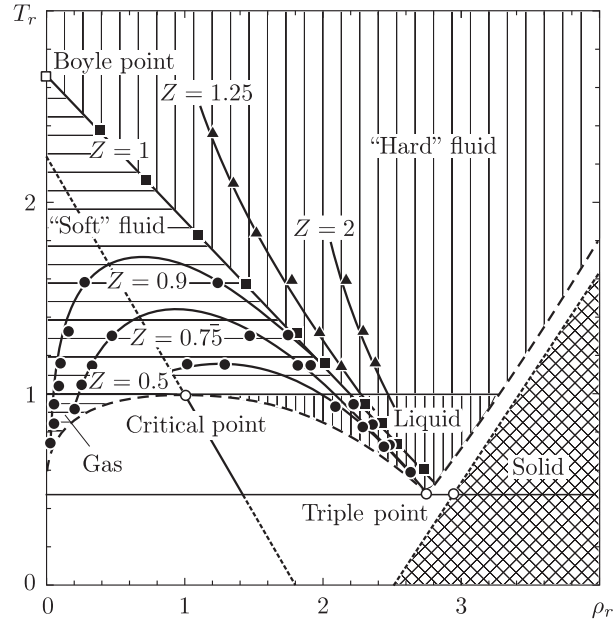}
\caption{$T$--$\rho$ diagram for gases corresponding to simple liquids,
$T_{r}=T/T_{cr}$, $\rho_{r}=\rho/\rho_{cr}$. The $Z=\frac{PV}{NT}=1.0$ line
(Zeno--line) on the phase diagram.For states with $Z>1.0$ (hard fluids)
repulsive forces dominate. For states where $Z<1.0$ (soft fluids) attractive
forces dominate.}\label{fig21}
\end{center}
\end{figure}

\bigskip

Further, for a fixed $\rho$, which is proportional to $\alpha$, we obtain the
asymptotic behavior
\[
E(r)=\frac{r^{2}\Phi(r)-\rho r^{4}}{B^{2}-r^{2}},\qquad\text{as}\quad
B\rightarrow\infty,
\]
where $B$ is the impact parameter, and also the ratio of the difference of the
maximal and minimal points of $E(r)$ to the maximal point. By our
correspondence principle, this ratio corresponds to the compressibility
factor
\[
Z=\frac{E_{\max}-E_{\min}}{E_{\max}}
\]
and, as $B\rightarrow\infty$, for a given $\rho$, we obtain the minimum value
of $Z$ on the graph $(\rho,T)$. The Zeno line has already been obtained by the
rule given above.

The value of the compressibility factor $Z$ is already plotted along the $y$
axis. We must now establish the correspondence with the temperature scale. To
do this, consider the ordinate axis, i.e., the case $\rho=0$. The point
$0.8\varepsilon$, where $\varepsilon$ is the depth of the well of the
Lennard-Jones potential corresponds to the Boyle temperature.

The minimal value of the compressibility factor $Z$ for a given  $\rho$
is equal to
\begin{equation}\label{Zmin}
Z_{\min}(\rho)=\frac{E_{\min}}{E_{\max}}\bigg|_{B=B_{\max}}.
\end{equation}
In this problem, eliminating the parameter $B$,
we found the Zeno line from the condition  $E_{\max}=E_{\min}$, and, as
$B\to\infty$ and for $C_2=1$,  we see that
\begin{equation}\label{d2}
Z=1-\frac{E_{\min}}{E_{\max}}\bigg|_{B\to\infty}.
\end{equation}
This gives a curve shown in Fig.~2 on the plane $Z,\rho$
or the curve presented in the Fig.~3 in the coordinates $\rho/\rho_B$.

\begin{figure}
\begin{center}
\includegraphics{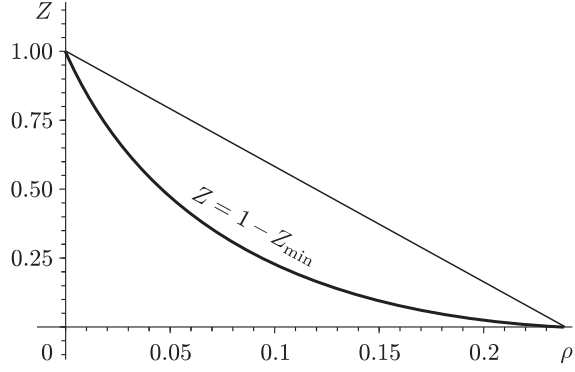}
\caption{The curve  $Z=1-Z_{\min}$ for $C_2=1$.} \label{fig2}
\end{center}
\end{figure}

\begin{figure}
\begin{center}
\includegraphics{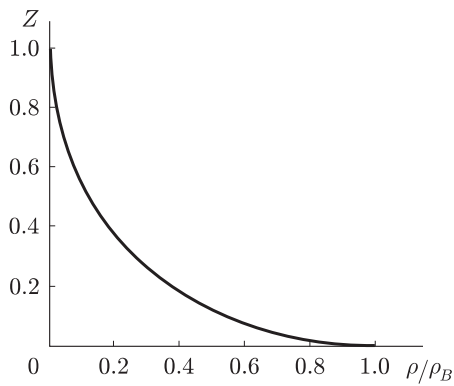}
\caption{The curve $Z=1-Z_{\min}$ in the coordinates $\rho/\rho_B$.} \label{fig3}
\end{center}
\end{figure}

We note that the values $Z$ of formula~\eqref{d2} for $B\geq 10$ remain almost unchanged,
i.e., $B=10$ is a large parameter.

Since for $Z<Z_{cr}=0.29$ (the point at which the derivative
along the diagonal in Fig.~3 vanishes), we arrive
at a contradiction related to two large parameters  $B$ and  $V$,
these parameters are to be coordinated  with one another.

For $C_2=0$  we consider the pair interaction only.
The  M-tame hyperbola corresponds to this ``jump''.
Further near the point  $Z=0.444$  there must be a short M-wild curve,
joined to a new M-tame curve, for which $C_2=0$. This curve is related
to the dimensionless parameter $V^{-1/3}B$.

Let us find a point
$$
Z=0.444=\frac 32 Z_{cr}.
$$
At this point we have
$$
V=\frac 32 V_{cr},
$$
if the reduced coordinates
$$
T_r=\frac{T}{T_{cr}}, \qquad P_r=\frac{P}{P_{cr}}
$$
are considered.
The value of  $B$  at the point  $1-Z$ is equal to $2.271$.
The critical point $Z_{cr}, \rho_{cr}$ is connected to this point by a hyperbola,
see Fig.~5.

At the point $\rho=0.19$ there is a modification of the curve of a rather large
scale which we refer to as the passage from an $M-tame$ curve to an  $M-wild$  curve,
where $M$ stands for the scale.  We had expressed a conjecture that this point
is related to the so-called triple point.

After studying the supermodern experimental data,
the author resigns the idea that a pressure jump happens simultaneously
to the volume jump. The chaos of clusters which happens
in the entire conception of the author is not an obstacle to this resignation.
The binodals which were constructed by the author earlier  correspond to the
nonequilibrium states and an attempt to find a pressure jump with
the accuracy up to which thermodynamical problem is solved turned out to be futile.

A thorough analysis of the modern experiments concerning
the measurements of the pressure of saturated vapor showed that,
when at least somewhat receding from the critical point, the pressure
is equal to the pressure inside the liquid, independently of Earth's attraction
which was mentioned above. Indeed, at the expense of thermal Brownian  motion,
when establishing the equilibrium at large time, the pressure  become equal indeed with
sufficiently large accuracy.

In the table, we present the data corresponding to the resulting diagram (for
$B=100$ in ``molecular'' units) and note the
disagreement between the main dimensionless relations resulting from the data
from molecular dynamics and the theoretical relations obtained by physicists
from the the BBGKY hierarchy of equations and the $n$-particle Gibbs distribution.

\begin{table}[ht]
\caption{}
$$
\begin{tabular}
[c]{ccc}
$Z_{cr}$ & $\rho_{cr}/\rho_{B}$ & $T_{cr}/T_{B}$\\
0.29 & 0.273 & 0.39\\
0.308 & 0.285 & 0.38\\
0.375 & 0.333 & 0.296
\end{tabular}
$$
\end{table}

The upper row of the table contains the theoretical values of $Z_{cr},$
$\rho_{cr}/\rho_{B},$ and $T_{cr}/T_{B}$ obtained by the author, the second
row contains the values of these quantities resulting from the newest data of
molecular dynamics for the Lennard-Jones potential, and the third row gives
the values based on the Van der Waals equation, which is empirical and even
``more'' empirical than the law of
correspondence given in this section.

For the ratio $T_{cr}/T_{B}$, different data are given in different reference
books; this is due to the fact that it is very difficult to determine $T_{B}$:
molecular dynamics does not provide the value of~$T_{B}$. In the physical
literature, the interaction potential for inert gases is given; this is the
Lennard-Jones potential. Therefore, it is interesting to compare the data
obtained on the basis of the correspondence principle with the corresponding
values for gases.

The value of $Z_{cr}$ was experimentally determined with a high degree of
accuracy and it is $0.29$ for inert gases, nitrogen, oxygen, and propane.

In the correspondence principle given by the author, the value of $\rho
_{cr}/\rho_{B}$ (the ratio of the critical $\rho$ to $\rho_{B}$, i.e., to the
whole length of the interval of $\rho$ values, where the Zeno line
``cuts off'' the abscissa axis on the graph
in Fig.~2 coincides with the corresponding values for water, argon, xenon,
krypton, ethylene, and a number of other gases.

Since the physicists (with the exception of V.~L.~Ginzburg) did not believe
the heuristic considerations that led to this rule of correspondence, let me
present detailed calculations to determine the compressibility factor and the
analog of the Zeno line for an interaction potential $U(r)$ of either of the
following types - generalized Lennard-Jones, Buckingham, Kohara, Morse,
Schommers, Barker.

The potential energy has the form
\begin{equation}
E(r)=\frac{-\alpha r^{4}+r^{2}U(r)}{B^{2}-r^{2}}. \label{ch1}
\end{equation}
The first derivative is
\begin{equation}
E^{\prime}(r)=\frac{2B^{2}rU(r)+2\alpha r^{3}(r^{2}-2B^{2})+r^{2}(B^{2}
-r^{2})U^{\prime}(r)}{(B^{2}-r^{2})^{2}}. \label{ch2}
\end{equation}
The second derivative is
\begin{align}
E^{\prime\prime}(r)  &  =\frac{1}{(B^{2}-r^{2})^{3}}[2(B^{4}+3B^{2}
r^{2})U(r)-2\alpha r^{2}(6B^{4}-3B^{2}r^{2}+r^{4})\label{ch3}\\
&  +4r(B^{4}-B^{2}r^{2})U^{\prime}(r)+r(B^{2}-r^{2})^{2}U"(r)].\nonumber
\end{align}

The first derivative $E^{\prime}(r)$ is equal to zero if
\begin{equation}
\alpha=\frac{-2B^{2}U(r)-B^{2}rU^{\prime}+r^{3}U^{\prime}(r)}{2r^{2}
(r^{2}-2B^{2})}. \label{ch4}
\end{equation}
The second derivative $E^{\prime\prime}(r)$ is equal to zero for
\begin{equation}
\alpha=\frac{2(B^{4}+3B^{2}r^{2})U(r)+4r(B^{4}-B^{2}r^{2})U^{\prime}
(r)+r^{2}(B^{2}-r^{2})^{2}U"(r)}{2r^{2}(6B^{4}-3B^{2}r^{2}+r^{4}).}.
\label{ch5}
\end{equation}
In order to define the Zeno line, we equate $\alpha$ standing in relation
(\ref{ch4}), (\ref{ch5}):
\begin{align*}
&  \frac{-2B^{2}U(r)-B^{2}rU^{\prime}+r^{3}U^{\prime}(r)}{2r^{2}(r^{2}
-2B^{2})}\\
&  =\frac{2(B^{4}+3B^{2}r^{2})U(r)+4r(B^{4}-B^{2}r^{2})U^{\prime}
(r)+r^{2}(B^{2}-r^{2})^{2}U"(r)}{2r^{2}(6B^{4}-3B^{2}r^{2}+r^{4}).}.
\end{align*}
This results in the following equation:
\begin{equation}
-8B^{2}U(r)+2B^{2}rU^{\prime}(r)+r^{3}U^{\prime}(r)+2B^{2}r^{2}U"(r)-r^{4}
U"(r)=0. \label{ch6}
\end{equation}
Substituting the solution of this equation into relations (\ref{ch1}) and
(\ref{ch4}), we obtain (in the parametric form) curve $E(\alpha)$ that, in
case of the Lennard-Jones potential, is represented by the analog of the Zeno line.

Further, setting $E^{\prime}(r)$ to be equal to zero and taking $B$ to be
sufficiently large ($B=100$), we get two values of $r.$ Denote by $E_{\max}$
and $E_{\min}$, where $E_{\max}$ $>E_{\min}$, the values of $E(r)$
corresponding to these $r$ and consider $Z=(E_{\max}-E_{\min})/E_{\max}$. We
determine $T_{cr}(Z,\rho/\rho_{cr})=T_{cr}=1$ in reduced coordinates by using
graphs in Fig.~1 for $T=T_{cr}$ at the point of maximum of the binodal
$Z=Z_{cr}$ and for $Z\geq Z_{cr}$.

\begin{figure}[ht]
\begin{center}
\includegraphics{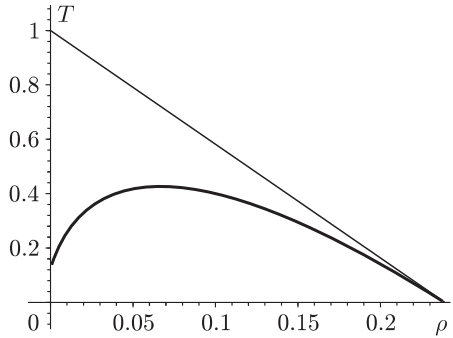}
\caption{The binodal and the Zeno line.} \label{fig5}
\end{center}
\end{figure}

\section{Constraint constants}

Van der Waals wrote his remarkable equation of state with two defining
constants $a$ and $b$. However, the Lennard-Jones potential describing the
interaction between particles (6--12) and also containing two defining
constants does not lead to the Van der Waals equation either in theoretical
calculations~\cite{Maslov_Arx_2009} or in experimental data even for noble
(inert) gases.

Following our ideas based on the analogy with economic laws, we shall
construct equations of state with reference to \textit{three} fixed points.

First, consider the Irving Fisher economic law for assets
\begin{equation}\label{87_5_1}
PQ=Mv,
\end{equation}
where $PQ$ is the amount of merchandise in its money equivalent, ($Q$ is the
merchandise and $PQ$ is its money equivalent), $M$ is the money supply, and
$v$ is the velocity of circulation.

Suppose that $M_{0}$ is some original money supply. (If the Fisher law is
considered historically, then it is expressed via its gold equivalent.) Let us
consider constraints on formula (\ref{87_5_1}) related to the overproduction of
merchandise in the form of the following statement. There exist constants~$a$
and~$b$ such that, for
\[
PQ\geq bM_{0}\qquad\text{or}\qquad v\geq c,
\]
there is no money supply~$M$ satisfying relation (\ref{87_5_1}). More
precisely, the money supply~$M$ satisfying relation~(\ref{87_5_1}) exists only
under two conditions:
\begin{equation}
PQ\leq bM_{0},\qquad v\leq c. \label{87_5_2}
\end{equation}
The constraint~(\ref{87_5_2}) is similar to the restriction on the velocity of
particles in mechanics. As is well known, it led to the revision of the
Newton--Galileo mechanics.

Now consider the equation of an ideal gas
\begin{equation}
PV=NT, \label{87_5_3}
\end{equation}
where $P$ is the pressure, $V$ is the volume, $N$ is the number of particles,
and $T$ is the temperature. Suppose that $N_{0}$ is the number of particles as
$T\rightarrow0$, $P\rightarrow0$. Then a similar statement is of the form:
relation (\ref{87_5_3}) holds only under the condition
\[
T\geq c,\qquad E=PV\geq bN_{0},
\]
where $c$ and $b$ are some constants.

Relation (\ref{87_5_3}) for an ideal gas can be represented as
\begin{equation}
P=T\rho, \label{87_5_4}
\end{equation}
where $\rho=N/V$ is the density of the gas. Then our constraint can be written
as follows:

\textit{For a given gas, there exist constants $c$ and $b$ such that if $T>c$
or $P>b$, then there is no density $\rho$ satisfying relation} (\ref{87_5_3}).

This question can be stated differently. Suppose that the pressure $P$ and the
temperature $T$ are given. Does there exist a density~$\rho$ satisfying
relation (\ref{87_5_4}) for a given imperfect gas (or a mixture of gases)?

In modern theory of imperfect gases, it is usual to consider the plane
$\rho,T$ and, in this plane, condition (\ref{87_5_3}) corresponds to the
so-called Zeno-line, which, as given by experiment, is a segment of the
straight line
\begin{equation}
\rho=\rho_{0}\left(  1-\frac{T}{T_{B}}\right)  ,\qquad T<T_{B}, \label{87_5_5}
\end{equation}
where $g_{0}$, $T_{B}$ are constants (the constant $T_{B}$ is called the
``Boyle temperature'').

This dependence was first noticed by Bachinskii, who justified it empirically.
Let us present the modern general scheme for an imperfect gas in the plane
$T,\rho$.

As was already pointed out by the author \cite{Maslov_87_3}, the thermodynamic
equations of state containing the quantities: $P$, the pressure, $T$, the
temperature, $\mu$, the chemical potential and, respectively, $V$, the volume,
$S$, the entropy, $N$, the number of particles, constitute a three-dimensional
Lagrangian manifold in the six-dimensional phase space; moreover, $P,$ $T,$
$\mu$ play the role of coordinates, while $V$, $S$, $N$, respectively, play
the role of the corresponding momenta.

Therefore, for a fixed number of particles $N$, the diagram on the plane
$\rho=N/V,\,T$ given in Fig.~1 is the projection on one of the planes of the
phase space. Under such a projection, focal and caustic points appear.

The slanting line issuing from the Boyle point in Fig.~1 is called the Zeno
line and is of the form (\ref{87_5_5}).

In studying the pressure as a function of the density $\rho$, an ambiguity
arises on the Zeno line:
\begin{equation}
P=c\rho\left(  1-\frac{c\rho}{4b}\right)  . \label{87_5_6}
\end{equation}
Hence the quantity
\begin{equation}
\rho=\frac{2b}{c}+\frac{1}{c}\sqrt{4b^{2}-4bP} \label{87_5_7}
\end{equation}
takes two values for $P<b$ and becomes complex for $P>b$, a typical simple
caustic (or the turning point in the one-dimensional quantum Schr\"{o}dinger equation).

\section{Compression on the plane~$P,V$ in Bose--Einstein-type
distributions and the fractal dimensions}

First, let us describe the distribution for $Z\leq Z_{cr}$.

An analog of the potential $\Omega_{\gamma}$
for the number theoretical distribution $\gamma= (d-1)/2$,
where $d$ is a ``fractal'' non-integer dimension and $d=D/2$,
has the form
\begin{equation}\label{87_5_9a}
\Omega^{\text{id}}_{\gamma}=\bigg(\frac{\pi^{1+\gamma}T^{2+\gamma}}{\Gamma(2+\gamma)}
\times
\int_0^\infty\xi^{1+\gamma}\bigg\{\frac{1}{e^{(\xi-\kappa)}-1}
\bigg\}d\xi \bigg),\qquad
\kappa=\frac{\mu}{T}, \quad T=\frac{1}{\beta}.
\end{equation}

Our distribution contains multiplication by a function of $V$,
i.e., the following change occurs in the Bose--Einstein distribution:
\begin{equation}\label{bc}
V\rightarrow\varphi_{\gamma}(V), \qquad \frac{\varphi_{\gamma}(V)}{V} \to 1
\quad \text{as} \   V\to\infty.
\end{equation}
The index $\gamma$ is sometimes omitted, and hence here it is constant.
This function is constant for a given dimension.
Therefore, $Z_{\min}$ obtained for $\mu=0$ has the form
\begin{equation}\label{f6}
Z_{\min}=\frac{V\varphi_{\gamma}'(V)}{\varphi_{\gamma}(V)}\cdot\frac{\zeta(d+1)}{\zeta(d)}=0.29,
\end{equation}
where $\zeta$ is the Riemann function.

Let us introduce one more quantity.
The press $\mathbb{P}$ is an intensive quantity conjugate to
the fractal dimension $\gamma$
$$
\mathbb{P}=\frac{\partial\Omega\gamma}{\partial\gamma}.
$$

For any $Z<1$ we have
\begin{equation}\label{f7}
Z=\frac{V\varphi'_{\gamma}(V)}{\varphi_{\gamma}(V)}\cdot
\frac{\Gamma(\gamma+1)}{\Gamma(\gamma+2)}\cdot
\frac{\int_0^\infty\frac{\varepsilon^{\gamma+1}\,d\varepsilon}
{e^{\xi-\kappa}-1} } {\int_0^\infty\frac{\varepsilon^{\gamma}\,d\varepsilon}
{e^{\xi-\kappa}-1} }
=\frac{V\varphi'_{\gamma}(V)}{\varphi_{\gamma}(V)}\Psi(\kappa), \qquad
\kappa=\frac{\mu}{T}, \quad  \varphi'_{\gamma}(V)=\frac{\partial \varphi}{\partial V},
\end{equation}
where $\Gamma(\cdot)$ is the gamma function.
For $\kappa=0$, we obtain~\eqref{f6}. Further, we find $\mu(V)$
as a function of $V$ from the condition $Z=1$:
\begin{equation}\label{dist6}
\frac{V\varphi'_{\gamma}(V)}{\varphi_{\gamma}(V)}\Psi(\kappa)=1, \qquad \kappa=\kappa(V).
\end{equation}

On the other hand, on $Z=1$, we have the Bachinskii parabola,
which follows from the Zeno line.
\begin{equation}\label{dist7}
P=\rho_B T\big(1-\frac{T}{T_B}\big)
\end{equation}
or
\begin{equation}\label{dist8}
P=T_B\rho\big(1-\frac{\rho}{\rho_B}\big),
\end{equation}
as well as
\begin{equation}\label{dist9}
T=T_B\big(1-\frac{\rho}{\rho_B}\big).
\end{equation}

Hence the dependencies $P(T)$, $T(\rho)$, and $P(\rho)$ are known:
$P(T)$ is the Bachinskii parabola,
$T(\rho)$ is a straight line,
and $P(\rho)$ is a parabola.

Let us find $P$ according to the Bose--Einstein distribution
with $V$ replaced by $\varphi_{\gamma}(V)$,
\begin{equation}\label{f8}
P=\frac{\varphi'_{\gamma}(V)T^{\gamma+2}}{\Gamma(\gamma+2)}
\int^\infty_0\frac{\varepsilon^{\gamma+1}\,d\varepsilon}
{e^{-\kappa}e^{\varepsilon}-1}.
\end{equation}

On the Zeno line~\eqref{dist7}, substituting~\eqref{dist8} and \eqref{dist9},
we obtain the second relation for $\varphi'_{\gamma}(V)$
under the additional initial condition that $\frac{\varphi_{\gamma}(V)}{V}\to 1$
as $V\to\infty$. The first-order differential equation
is obtained in parametric form (here $\kappa$ is a parameter).

The function $\varphi_{\gamma}(V)$ allows us to determine
the critical dimension $d$, and hence $\gamma_0=0.2$
as follows from~\eqref{f6} with $V=V_{cr}$.
Therefore, $\varphi_{\gamma_0}(V)$ gives us the solid curve in Fig.~5,
which is in a good agreement with experimental values for the argon gas component
(for $P\leq 1$).

\bigskip
\noindent\textbf{Example 3. The curve $T_r=1$
for ideal and imperfect Bose gas in $Z,\, P$ coordinates.}

The formula for ideal gas is
$$
\frac{Z}{P}{Li_{\gamma_0+1}(a)}=\zeta(\gamma_0+2),
$$
where $a=a(P)$ is determined by the equation
$$
\frac{Li_{\gamma_0+2}(a(P))}{\zeta(\gamma_0+2)}=P.
$$

The formula for imperfect gas is
$$
f\big(\frac{Z}{P}\big){Li_{\gamma_0+1}(a)}=\zeta(\gamma_0+2),
$$
where $a=a(Z,P)$ is determined by the equation
$$
\frac{Li_{\gamma_0+2}(a)}{\varphi'_{\gamma_0}(V_{cr})\zeta(\gamma_0+2)}
=\frac{P}{\varphi'_{\gamma_0}(\frac ZP)}.
$$
The results of computations by the last formula coincide
with experimental graph in Fig.~6
for $T_r=1$ up to the point $Z=0.29$.

\begin{figure}
\begin{center}
\includegraphics{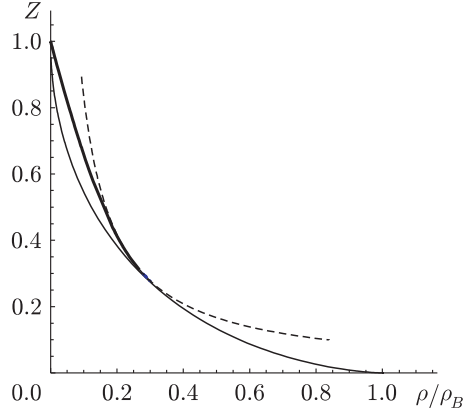}
\caption{The curve $Z(\rho)$ goes from the point $Z=0.1$ to the point
$Z_{cr}= 0.29$ along the thin curve, after which goes along the thick curve.
The  hyperbola is dotted.} \label{fig4}
\end{center}
\end{figure}

\begin{figure}[t]
\begin{center}
\includegraphics[height=3.0753in,width=4.3734in]{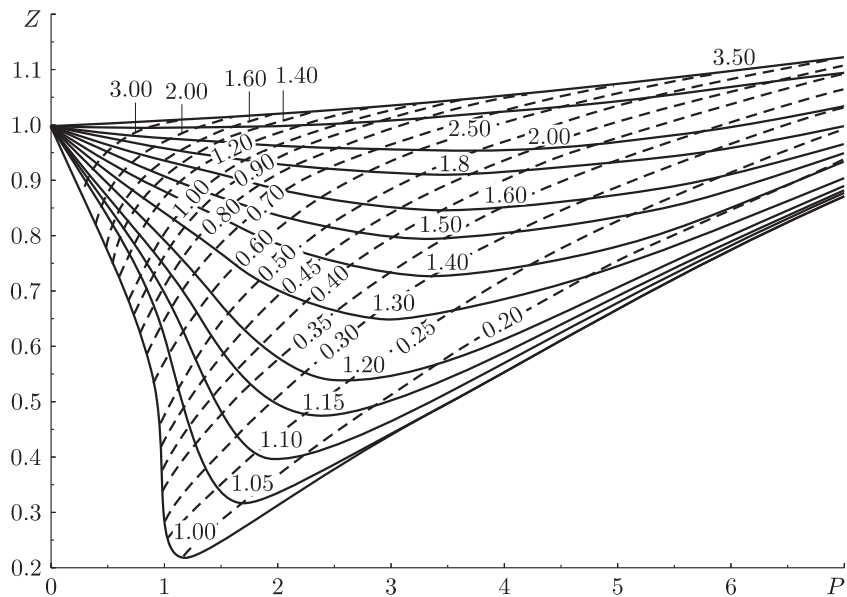}
\caption{Experimental graph.
$P_{r}=P_{atm}/P_{cr}$, $T_{r}=T/T_{cr}$,
$T$ is the temperature in Kelvin degrees,
$V/R$ is the volume in $cm^{3}/mole$,
$R$ is the gas constant, and
$Z=P_{r}V/(RT_{r})$ is the compressibility factor.
The isochores $V/R=const$ are shown by dotted lines.}
\end{center}
\end{figure}
\medskip

How  does the curve~$T_r=1$  go  below the  point  $Z=0.29$?

The compressibility factor $Z$ in this distribution is determined as the
following relation
\begin{equation}\label{JF_16a}
Z=\frac{P_r V}{T_r},
\end{equation}
where $P_r=P/P_{cr}$,  $T_r=T/T_{cr}$.
Hence, the fractal dimension $d$ is uniquely determined for $Z=Z_{cr}=V_{cr}$.

If we assume that the parameters $\beta=1/T$, $\mu$, $\kappa=\mu/T$, and $\gamma$
are mutually related by the condition that the number $L$ of versions
of the solution of the Diophantine equations for ${\mathcal{E}}$ and $N$
must increase maximally, then this means that the ``specific entropy'' $S$,
i.e., the entropy used in number theory , takes the maximum values.
Since
\begin{equation}\label{new63}
S_{id}=Z^{id}-\kappa\to\max, \qquad
Z^{id}_{\gamma+1}=\frac{Li_{\gamma+2}(e^\kappa)}{Li_{\gamma+1}(e^\kappa)},
\end{equation}
where $Li$ is a polylogarithm.

The condition that there is a relation between $\mu$ and $\gamma$
completely determines the curve $T_r=1$ in Fig.~6.
The same holds for the curves $T_r=\text{const} >1$.
We write the system of equations
\begin{equation}\label{new64a}
\dot\mu=\partial S_{id}/\partial\mu, \qquad
\dot\gamma=\partial S_{id}/\partial\gamma.
\end{equation}
The limit point is determined by the condition
$Li_{\gamma+1}(e^\kappa)=\text{const}$:
\begin{equation}\label{new64}
\frac{d}{d\mu}(Z^{id}_{\gamma+1}-\kappa)=0,
\end{equation}
and the constant is determined by the relations on the Zeno line (see Fig.~8).
The other boundary conditions are determined by the curve in Fig.~3
for $Z> Z_{cr}$.

Using the dependence $Z(\rho)$ ($\rho=1/V$)
on the thin curve for $Z>0.29$ (we mean that, on this curve, $\kappa$ is zero)
and the dependence on the Zeno line,
we  obtain that for $\gamma\geq\gamma_0$ the variables
$\varphi_{\gamma}(V)$  depend on $\gamma$
without a boundary condition of the type~\eqref{bc}
(for $\gamma\leq\gamma_0$  the variables $\varphi_{\gamma}(V)$
do not depend on $\gamma$). The  boundary condition is adjusted so that
the above relations and the continuity condition are satisfied.

From~\eqref{new64a} we have
\begin{equation}\label{new65}
\frac{d\gamma}{d\mu} =-TZ_{\gamma}^{id}\frac{\partial\ln Z_{\gamma+1}^{id}(e^{\kappa})}{\partial\gamma},
\qquad  \gamma=0.2 \quad  \text{as} \  \mu=0, \qquad T_r=1.
\end{equation}

We draw a line between  the breakpoint~$\gamma_1(\mu)$ and   the Zeno line   at the point $T_r=1$
($P=\rho_B(1-1/T_B)$, $Z=1$)(jamming effect). The variables   $\varphi'_{\gamma_1}$  are practically equal to constant.

By putting $\gamma(\mu)$ in
$$
P=\frac{\varphi'_{\gamma(\mu)}(V)Li_{\gamma(\mu)+2}(e^{\mu})}
{\varphi'_{\gamma_0}(V_{cr})\zeta(\gamma_0+2)}
\qquad\text{and}\qquad
Z=\frac{\varphi'_{\gamma(\mu)}(V)V}{\varphi_{\gamma(\mu)}(V)}\frac{Li_{\gamma(\mu)+2}(e^{\mu})}
{Li_{\gamma(\mu)+1}(e^{\mu})},
$$
we obtain the  extension of the curve~$T_r=1$ corresponding to Fig. 6.

Since it turns out that $\gamma'$ is close to unity,
we present a figure, where $\gamma$ linearly depends on $V$,
and compare it with the experimental values on the plane $Z,P$.
The character of the behavior of $\gamma(V)$
and the fluid incompressibility effect as $\gamma\to0$ (the jamming effect)
can already be seen in this approximation (Fig.~7).

For $Z<0.17$, the jamming effect already takes place, and this permits improve
the binodal, i.e., the phase equilibrium curve. For $\gamma<\gamma_0$,
the function $\varphi_{\gamma_0}(V)$ is obtained, and hence all quantities
determined by the potential $\Omega_{\gamma}$, can also be determined.

\begin{figure}[ht]
\begin{center}
\includegraphics[height=3.6469in,width=4.3889in]{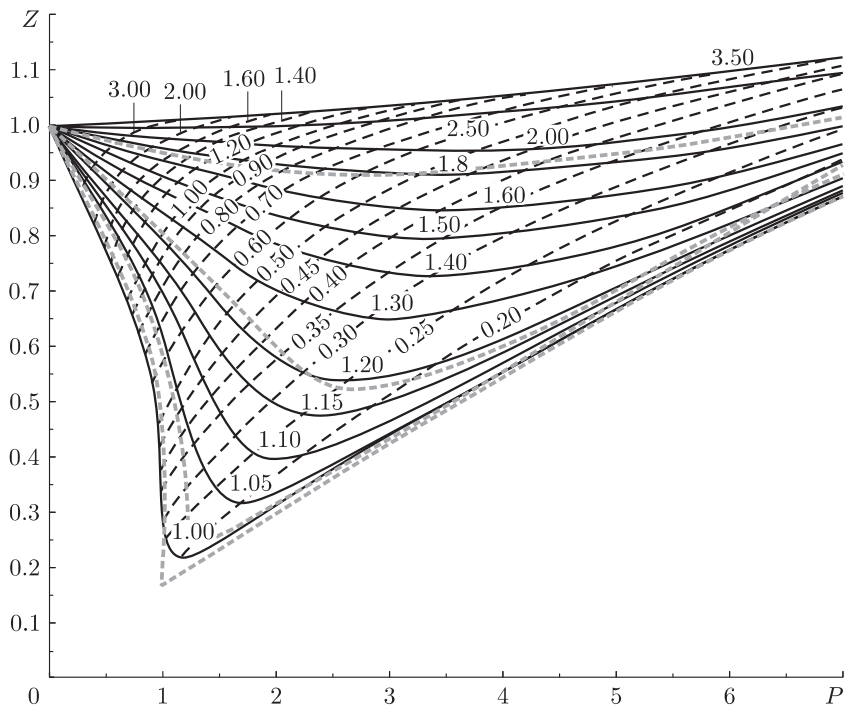}
\caption{The isotherms for $T_r>1$.
The bold lines denote theoretical isotherms for $\gamma'=1$.}
\end{center}
\end{figure}
\medskip

On the phase equilibrium curve, we know the gas pressure and temperature.
They are equal to the pressure and temperature of the liquid phase.
For $Z< 0.17$, the picture shown in Fig.~8 takes place,
where $\rho$ and $T$ do not vary along the rays and are determined on the Zeno line.

\begin{figure}[ht]
\begin{center}
\includegraphics{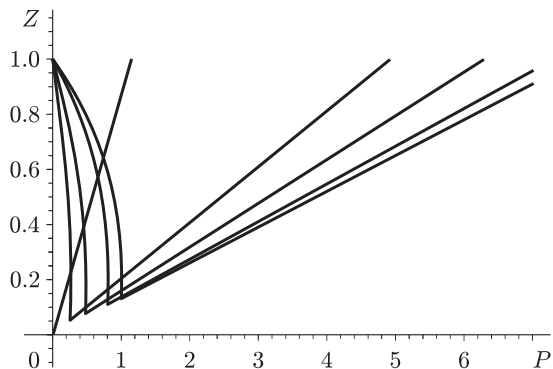}
\caption{Isotherms in the liquid phase region.
For $Z<0.17$, formula degenerates, and the rays $T=const$
stick into the Zeno line according
to formula~\eqref{dist9}--\eqref{new63}.
It follows from condition \eqref{new63} that, for $\mu<0$,
the relation between $\mu$ and $\gamma$ are such that
$Li_{\gamma+1}(e^\kappa)=const$ for a fixed~$T$.
Condition \eqref{new64} with (11) for $k=N$ taken into account
implies a shift in $\mu$ of order $1/\log_2 N$.
This transition on an enlarged scale can be obtained from \eqref{new64} and (11).}
\label{fig_8}
\end{center}
\end{figure}

Since, for the gas branch, we determined the value of~$T$
in the triple point, the corresponding value of $\rho$
for the liquid phase we obtain from the equilibrium condition.
Thus, we obtain $Z_{cr}^{triple}\approx 0.3\cdot 10^{-3}$,
$T_{cr}^{triple} \sim 0.55\, T_{cr}$,
$\rho_{cr}^{triple} \approx 0.7 \text{gm}/\text{cm}^3$.

This allows us to introduce one more critical point for the liquid phase
and to find $\varphi_{cr}^{triple}(V)$, just as this was done
in the case of gas phase, which means that a new potential $\Omega_{\gamma_{triple}}$
can be introduced in the liquid phase.
This is an important fact, useful in liquid dynamics (see~\cite{MZ_88_6}).

Since, in this domain, $\rho$ and $Z<Z_{cr}$ are related by $Z\cong c/\rho$,
where $c=const$, it follows that the decrease in the fractal dimension is
determined by the increase in the value of density~$\rho$, and hence also by
the increase in the pressure. The constant~$c$ is determined by the condition
$c=Z_{cr}\rho_{cr}$. The increase in the pressure can be calculated by the
successive approximation method.

In addition, our distribution must be consistent with the Zeno line, which is
valid, according to experiments, for a wide range of spectra of different
gases. Let us show this using an example (given below). However, the main
thing is that, in our distribution, the dimension is a given function of the
density $\rho$.\medskip

Below the condensate point, for $P=1$, the angle of rotation depends on
$V_{\gamma}/R$ for $V/R=Z<Z_{cr}=0.3$. This is a fairly complicated
transformation. Nevertheless, since it occurs in the plane $\widetilde{P},$
$\widetilde{V}/R$, it does not involve the coordinates $T,$ $S/R$ and
preserves the Lagrange property, because, in the two-dimensional phase space,
any smooth transformation preserves the Lagrange property and the coordinates
$T,$ $S/R$ remain unchanged. It is important that $S$ is still the logarithm
of the number of possible variants.

Let us now pass to the projection of the two-dimensional manifold (obtained
above) in the six-dimensional space on the four-dimensional phase space. The
equality $N=const$ cuts a two-dimensional surface out of the three-dimensional
Lagrangian manifold. The following group property holds: as $N$ is changed
$k$-fold, so are the quantities~$V$ and~$S$. This implies that this manifold
is cylindrical. Therefore, it can be projected along~$\mu$ and~$N$ from the
six-dimensional phase space on the four-dimensional space, where the explicit
action determining the Lagrangian manifold is the Gibbs potential
$\widetilde{\Phi}=N\mu(P,T)$.

Since we have $N=const=R$, this equality cuts out a two-dimensional
cylindrical (by the group property) manifold depending on the parameter
$d=2\gamma+2$. Projecting it along~$N$ and~$\mu$, we obtain a two-dimensional
manifold in the four-dimensional phase space~$P,V/R,T,S/R$. First, consider
this projection using $\Omega_{\gamma}^{id}$ as an example.

Here $P,$ $T$ play the role of coordinates, while $V/R,$ $S/R$ that of
momenta. Using relation $P=\partial\Omega/\partial V$, we obtain
\begin{equation}
P=\varphi'_\gamma T^{2+\gamma}\int_{0}^{\infty}\frac{\varepsilon^{1+\gamma}\,d\varepsilon
}{e^{-\mu/T}e^{\varepsilon}-1},\qquad\text{for}\quad\gamma>0 \label{JF_20}
\end{equation}
and
\begin{equation}
P=\varphi'_\gamma T^{2+\gamma}\int_{0}^{\infty}\xi^{1+\gamma}\left\{  \frac{1}{e^{(\xi
-\kappa)}-1}-\frac{N}{e^{(\xi-\kappa)N}-1}\right\}  \,d\xi,\qquad
\text{for}\quad1<\gamma\leq0. \label{JF_21}
\end{equation}

From (\ref{JF_20}) expressing $\mu$ as a function of~$P$ and $T$:
$\mu=\mu(P,T)$, we obtain the potential
\[
\widetilde{\Phi}=R\mu(P,T)
\]
and relations
\[
V=\frac{\partial\widetilde{\Phi}}{\partial P},\qquad S=\frac{\partial
\widetilde{\Phi}}{\partial T}\,.
\]
This gives rise to a potential of the type of the Gibbs potential $R\mu(P,T)$
determining the Lagrangian manifold non-uniquely projected onto the plane $P,$
$T$.

The usual rule is used for the selection of points at which different branches
of the projections are joined. The ``quantization'' of thermodynamics is carried out by the tunnel
canonical operator, as was described in \cite{Maslov_Kvantovan_book}.

Just as above, we can project the Lagrangian manifold (after the rotation) in
transformed (with respect to $V$ and $P$) coordinates along $\mu$ and $N$ on
the four-dimensional phase space. Since the projection does not depend on the
order of transformation of the coordinates $P$ and $V$, it follows that the
same transformation can be performed after the projection.

The fact that the potential ``mixed'' from
the free energy and the thermodynamic potential arises can easily be
explained; see the heuristic example given in \cite{Maslov_Arx_2009}. The
phase transition in liquids is treated there as the creation of
``three-dimensional'' clusters or domains, as
the author called them, in which there exists a particle surrounded on all
sides by other particles (constituting a small coordination sphere).

\begin{remark}\rm
The angles of rotation $\alpha$ (in radians) in the plane $P,V$ depending on
$V$ are given in the following table:

\begin{center}
\begin{tabular}[c]{cc}
$V\geq0.3$ & $\alpha=0.049$\\
$V=0.25$ & $\alpha=0.052$\\
$V=0.20$ & $\alpha=0.058$\\
$V=0.17$ & $\alpha=0.066$
\end{tabular}
\end{center}

\end{remark}

It should be noted that $V=0.17$ corresponds to the last limiting point to
which all the values $V>0.17$ accumulate. This point is a focus; therefore,
the ``quantization'' of thermodynamics by the
tunnel canonical operator \cite{Maslov_Kvantovan_book} strongly erodes this point.

Point $Z=0.17$ is a focal point, which causes the asymptotics to spread and
become smooth.

\begin{remark}\rm
In the final formula for the distribution of number-theoretic form, we must
also take into account the dependence on~$r$ that corresponds to the expansion
of the thermal potential $\Psi(r)$ up to the second power of $r$ in
(\ref{JF_5}).
\end{remark}

The phenomenon demonstrated by the graph in Fig.~7 is called \textit{jamming}.
It leads to the rapid incompressibility of a fluid, i.e., to the formation of
a glass dust, not a crystal (just as in a recent eruption of the Icelandic
volcano), which is almost insensitive to the pressure, i.e., becomes
incompressible. For $P>1.5P_{cr}$ and at the critical temperature $T_{r}=1$
$(T_{reduced}=T/T_{cr}$, and hence $Z=P/\rho$), line $Z=c+dP$ on the graph
$(P,Z)$ implies that $\rho=const$, despite the increase in pressure. This
``new'' condensate is not reflected by the
diagram $T-\rho$ in Fig.~1, where the transition to crystal is shown.

The explanation given to this phenomenon by physicists using the model of hard
balls is similar to the example of a two-dimensional billiard studied by the
author in~\cite{Maslov_ArXiv_Thresh_levels}. Here, as is readily verified, the
fractal dimension tends to zero along this line, and hence, near this
``new condensate,'' it is necessary to use
the global asymptotics of number theory presented by the author
in~\cite{Maslov_ArXiv_Thresh_levels},~\cite{MZ_87_5}.

From our above discussion it follows that for the construction of liquid phase
temperature curves $T_{r}=const$ on the graph $(P,Z)$, one has to use the
global asymptotics presented in Section~3. We note that, in general, in
addition to quantities considered in Section~3, the introduced approach allows
to determine also the entropy and the fractal dimension.

The energy spent on the increase in pressure is used not to increase the
density, but, rather, to effect an internal modification, meaning the decrease
in the fractal dimension, and hence the decrease in the number of degrees of
freedom. In \cite{Maslov_87_3}, the author called a graph of the type in Fig.~7
as a ``pit'' for the case of economic and
revolutionary crises when the revolution in the cause of ``freedom'' leads, as a result of a ``new
condensate'' to the successive decrease in the number of
degrees of freedom.

\section{$N$-particle Gibbs distribution containing all interactions}

Consider a distribution, averaged over different trials, whose number $L$ is
much greater than~$N$, of systems of $N$ particles at the same temperature
(mean energy), and it is also a distribution over energy surfaces
$H(p,q)=const$, $p\in\mathbb{R}^{3N}$, $q\in\mathbb{R}^{3N}$ (i.e., $N=const$,
$L\rightarrow\infty)$.

This distribution is a distribution over energy levels
\[
E_{1}\leq H(p,q)\leq E_{2}.
\]
Let us pass to the statement of the theorem.

Consider the Weyl quantized self-adjoint operator $\widehat{H}=H(\widehat
{p},q)$ with the discrete positive spectrum
\[
\text{ $\lambda_{1}\leq\lambda_{2}\leq\dots\leq\lambda_{n}\dots$
,\qquad$\lambda_{n}\rightarrow\infty$ \quad\text{as}\ \ $n\rightarrow\infty$ }
\]
and the corresponding proper subspaces $P_{1},$ $P_{2},$ $\dots,$ $P_{n}
,\dots$ in the Hilbert space $L_{2}(\mathbb{R}^{3N})$. We assume that the
number of eigenvalues less than a given~$E$ obeys Weyl's rule as
$E\rightarrow\infty$ (in physics, this is expressed as follows: the number of
eigenvalues is proportional to the phase volume). Sometimes this formula is
called Courant's formula.

Consider the tensor product of $L$ Hilbert spaces
\begin{equation}
{\mathcal{L}}_{2}=\underset{L}{\underbrace{L_{2}(\mathbb{R}^{3N})\otimes
L_{2}(\mathbb{R}^{3N})\otimes L_{2}(\mathbb{R}^{3N})\dots\otimes
L_{2}(\mathbb{R}^{3N})}}, \label{JF_22}
\end{equation}
where $L\rightarrow\infty$, and define the operator
\begin{equation}
\widehat{H}_{L}=\text{ }\underset{L}{\underbrace{\widehat{H}\otimes
1\otimes1\dotsm\otimes1+1\otimes\widehat{H}\otimes1\dotsm\otimes
1+\dotsm+1\otimes1\dotsm1\otimes\widehat{H}}}. \label{JF_23}
\end{equation}

Determine the parameter $b_{E}$ from condition
\begin{equation}
\frac{\mathrm{Sp}(\text{ }\widehat{H}e^{-b_{E}\widehat{H}})}{\mathrm{Sp}
(e^{-b_{E}\widehat{H}})}=E. \label{JF_24}
\end{equation}

Denote $L_{0}=\mathrm{Sp}(e^{-b_{E}\widehat{H}})$. The occupation numbers of
the operator $\widehat{H}_{L}$ corresponding to the eigenfunction $\Psi_{i}$
of the operator $\widehat{H}$ less than or equal to $EL$, are denoted
by~$L_{i}$. The sum $L_{i}+L_{i+1}+\dots+L_{i+k}$ is denoted by $L_{i}^{(k)}$,
the occupation numbers\footnote{To define the number $L_{i}^{(k)}$, whose
physical meaning is the number of particles, we can assume that the particles
are indistinguishable.} corresponding to the collection of functions $\Psi
_{i},$ $\Psi_{i+1},$ $\dots,$ $\Psi_{i+n}$, while the projection operator on
the subspace spanned by proper subspaces corresponding to them is denoted by
$P_{i}^{(k)}$.

For a Gibbs ensemble, the following theorem holds \cite{Maslov_TMFGibbs}.

\begin{theorem}
There exist constants $C_{l}$, $l=1,2,\dotsc$, such that, for all $i>1$ and
$n>0$, the occupation numbers $L_{i}^{(n)}$ satisfy the inequality
\begin{equation}
\mathsf{P}\left(  \left\vert L_{i}^{(n)}-B\mathrm{Sp}(P_{i}^{(n)}
e^{-b_{E}\widehat{H}})\right\vert >B\sqrt{L_{0}\ln L_{0}}\psi(L_{0})\right)
\leq C_{l}L_{0}^{-l},\text{ \ \ \ }l=1,2,\dotsc, \label{JF_25}
\end{equation}
where $B={L}/{L_{0}}$, $\psi(x)$ is a positive function tending arbitrarily
slowly to $+\infty$ as $x\rightarrow+\infty$, and $\mathsf{P}(\cdot)$ is the
ratio of the number of eigenvalues of the Hamiltonian $\widehat{H}_{L}$ which
do not exceed $LE$ and satisfy the inequality in the parentheses to the total
number of eigenvalues of the Hamiltonian $\widehat{H}_{L}$ which do not exceed
$LE$ (the spectrum density)\footnote{Landau and Lifshits used the term
``spectrum denseness'' ~\cite{Landau}, p.~44
for the microcanonical distribution.}.
\end{theorem}

Since $\sum\lambda_{i}L_{i}$ is the eigenvalue of the operator $\widehat
{H}_{L}$ corresponding to the eigenfunctions, it follows that the sum $\sum
L_{i}$ over all $L_{i}$ is the number of all eigenvalues of the operator
$\widehat{H}_{L}$ less than $LE$.

Theorem~1 implies that if we prescribe $L_{i}^{(n)}$ on the interval
$\lambda_{i},\dots,\lambda_{i+n}$ and calculate the ratio of the number of
eigenvalues with given $L_{i},$ $L_{i+1},$ $\dots,$ $L_{i+n}$ and other
arbitrary $L_{j}$ for $j<i$, $j>i+n$ to the total number of eigenvalues of the
Hamiltonian $\widehat{H}_{L}$ not exceeding~$LE$, then this ratio tends to
zero as $L_{0}^{-s}$ for any~$s$ outside the interval
\[
L_{i}^{(n)}\sim\mathrm{Sp}(P_{i}^{(n)}e^{-b_{E}\widehat{H}})\pm B\sqrt
{L_{0}\ln L_{0}}{\psi}(L_{0})
\]
for Theorem 1.

Here we are referring to the spectrum density of the Gibbs ensemble and to the
fact that formula (\ref{JF_24}) determines the mean value of the energy and
the value of~$LE$ bounding the spectrum density of the operator $\widehat
{H}_{L}$. We can also say that this quantity is the total number of
eigenvalues of the operator $\widehat{H}_{L}P_{LE}$, where $P_{LE}$ is the
projection operator onto all proper subspaces corresponding to all $\mu
_{n}^{(L)}\leq LE$, where the $\mu_{n}^{(L)}$ are all the eigenvalues of the
operator $\widehat{H}_{L}$ not exceeding $LE$.

Suppose that ${\mathcal{P}}_{i}^{n}$ is the projection operator onto all
proper subspaces of the operator $P_{LE}\widehat{H}_{L}$ lying outside the
interval enclosing the eigenvalues of the operator
\[
P_{i}^{n}\widehat{H}e^{-b_{E}\widehat{H}}\pm B\sqrt{L_{0}\ln L_{0}}\psi
(L_{0}).
\]
Then the ratio of the number of eigenvalues of the operator ${\mathcal{P}}
_{i}^{n}\widehat{H}_{L}$ to the number of eigenvalues of the operator
$P_{EN}\widehat{H}_{L}$ tends to zero faster than any power of~$L_{0}^{-1}$.

To pass to the classical Gibbs distribution, it is not necessary to pass to
the limit as $h\rightarrow0$. It suffices to define the self-adjoint operator
with eigenvalues equal to those intervals of the phase volume that are
satisfied by the eigenvalues. Then we only have to pass from the sums to
integrals by using the Euler--Maclaurin formulas.

Let us split the phase space $(p,q)\in\mathbb{R}^{6N}$ into a finite number of
domains
\begin{equation}
E_{l}\leq H(p,q)\leq E_{l+1}, \label{JF_24a}
\end{equation}
where $l=0,\dots,s-1$, $E_{0}=0$, $E_{s}=E$, $p\in\mathbb{R}^{3N}$,
$q\in\mathbb{R}^{3N}$, and $\mathbb{R}^{2NL}$ is the phase space with
coordinates $p_{1},q_{1},$ $p_{2},q_{2},$ $\dots,$ $p_{N},q_{N}$. Let us
perform an ordered sampling with replacement of $L_{l}$ from the partition of
the domains of the space $\mathbb{R}^{2NL}$ into the ``box'' $E_{l}\leq H(p,q)\leq E_{l+1}$, under the condition
\begin{equation}
\sum_{i=1}^{L}H(p_{i},q_{i})\leq LE. \label{JF_26}
\end{equation}

From the physical point of view, ordered sampling means that we consider $L$
distinguishable $3N$-dimensional particles. Let $\rho_{E_{l}}^{\Delta}$ denote
the cluster ``density'' in the energy
interval $E_{l}\leq H(p,q)\leq E_{l+\Delta}$.

Suppose that the above conditions on the function $H(p,q)$ hold. Let $L$ be
given. Let us determine~$b$ from the condition
\begin{equation}
\int_{0}^{\infty}e^{-bH(p,q)}\,dp\,dq=L,\qquad b=\frac{1}{kT}\,. \label{JF_27}
\end{equation}
We define~$E$ in (\ref{JF_26}) as
\[
\text{ $\int_{0}^{\infty}H(p,q)e^{-bH(p,q)}\,dp\,dq.$}
\]
Then the following theorem is valid.

\begin{theorem}
The following relation holds:
\begin{equation}
\mathsf{P}\left(  \left\vert L\rho_{E_{l}}^{\Delta}-\int_{_{E_{l}\leq
H(p,q)\leq E_{l+1}}}e^{-bH(p,q)}\,dp\,dq\right\vert \geq\sqrt{L\ln L}
\psi(L)\right)  \leq L^{-m}, \label{JF_28}
\end{equation}
where $m$ is any integer.
\end{theorem}

Here the probability $\mathsf{P}$ is the Lebesgue measure of the phase volume
given in parentheses (\ref{JF_28}) with respect to the whole phase volume
(\ref{JF_26}).

The proof of Theorem~2 is given in \cite{Maslov_ContMath}, Section~2; see also
\cite{Maslov_TMFGibbs}.

\section{On the Maxwell--Boltzmann distribution}

The distribution of the Bose-Einstein type distribution is determined from the
relation
\begin{gather}
\int_{0}^{\infty}\frac{H(p,q)\,dp\,dq}{e^{\beta(H(p,q)+\kappa)}-1}
={\mathcal{E}},\qquad p\in\mathbb{R}^{3},\quad q\in\mathbb{R}^{3},\nonumber\\
\int_{0}^{\infty}\left\{  \frac{1}{e^{\beta(H(p,q)+\kappa)}-1}\right\}
\,dp\,dq=\frac{N}{V}, \label{JF_29}
\end{gather}
where
\[
\beta=\frac{1}{T},\qquad\kappa=\frac{\mu}{T},\qquad H(p,q)=\frac{p^{2}}
{2m}+u(q).
\]
If $\mu\rightarrow-\infty$, we get the Maxwell--Boltzman distribution
$e^{-(p^{2}/2m+u(q))/T}.$

On the assumption that the potential field varies very slowly, the function of
the coordinates has the form
\[
u(q)=U\left(  \frac{q}{\sqrt[3]{V}}\right)  .
\]
Only on this assumption, we can pass from the Maxwell--Boltzmann distribution
$e^{-(p^{2}/2m+u(q))/T}$ (by integrating it over the momenta) to the Boltzmann
distribution of the form
\begin{equation}
e^{-\beta u(q)}, \label{JF_30}
\end{equation}
because the Maxwell--Boltzmann distribution is not a distribution of the
particle number density with respect to momenta and coordinates. It only gives
the number of particles between the energy levels
\[
\frac{p^{2}}{2m}+u(q)=const\text{.}
\]
Therefore, only if $u(q)$ is of the form
\[
u(q)=U\left(  \frac{q}{\sqrt[3]{V}}\right)  ,
\]
then this can be done in the thermodynamic limit.

Here we must not forget that the Boltzmann distribution is not also a
distribution with respect to coordinates $q$, but rather with respect to the
level surfaces of function $u(q)$.\bigskip

\section{Conclusions}

We began with usual statistics used in molecular physics
(see Section~3 before formula (24)).
We determined $Z_{cr}$ and Zeno line.
But \textit{only} Zeno line ($Z=1$) and the critical points are stable
under transition to statistics of identical particles (monomers).
The critical points give a very strong focus,
and $Z=1$ means the ``complete'' victory of Shannon's entropy.
We obtain a noble gas out of glassy dust
(the hydrodynamics of such a dust was studied in~\cite{Maslov_Kvantovan_book})
when most of its particles are distinguishable from each other.
Namely, in noble gases, there are no preferences for dimers, trimers, etc.
Therefore, thermodynamics of noble gases gives good models of the mixture
of Bose--Einstein statistics and Boltzmann statistics.
There is a remarkably simple law of the mixture of statistics here.
As $Z=1$, this law is called the Zeno line.

In the construction in which we define
the critical temperature, we consider
the difference of the energies of the stable
and unstable rest points. Furthermore,
we use only small friction and viscosity.
Indeed, as a result of small viscosity,
a classical particle,
after having flown ``just'' above the barrier,
will lose its energy and, on reflection,
will hit the barrier and
will continue hitting
the walls of the barrier and
the walls of the potential
in the form of a well
until it precipitates to the bottom.
(The value of the energy of the stable rest point is
$E_1$ and the value of the energy of the unstable rest point
$E_2$.)
 To knock out this particle
from the well trap, the required
kinetic energy must be ``slightly'' greater than
the difference
between the energies $E_1-E_2$ of the rest points.
This can interpreted
as a break-up of the dimer
in the collision
with
a fast monomer.
 The equilibrium
is violated
when
the depth $E_1-E_2$ of the well
decreases
and
its width increases
as the absolute value of the energy~$E_1$ of the stable rest point decreases
(as the impact parameter increases).
 It is natural to regard the increasing width
as a result of an increase
in the numbers of dimers in
the trap (in quantum theory, this
corresponds
to an increase in spectral density).  Dimers clusters can survive
and equilibrium
between the monomers
and the dimers can be preserved
only if the dimers themselves
combine into
cluster domains and create their own barrier,
which is a microanalog of a surface film.

In  problem~(30) as $C_2=0$,
different barriers
and wells
occur for different values of~$B$.
At the rest points,
$E_{\min}$ and $E_{\max}$,
the velocity is zero;
therefore, they
can be determined
only
from the potential term.

We are now dealing
not with just one particle, but
with a pair of particles
whose mass center is in the trap.
Therefore, the difference
$E_{\max} - E_{\min}$
is the energy needed
to knock out this pair
(the dimer) from the trap.

Experimentally,
we can calculate the percentage of dimers
in a gas.  It is clear how dimers are created
and split by monomers.  Further, their mean number is calculated.
The higher is the temperature,
the greater is the mean energy of monomers
and the smaller the number of dimers.

By our calculations,
we have $T_{cr}/T_B=2.79$.
According to the contemporary handbooks,
this value is $T_{cr}/T_B=2.72$
for argon (Ar),
$T_{cr}/T_B=2.71$
for krypton (Kr),
$T_{cr}/T_B=3.157$
for methane ($CH_4$),
and
$T_{cr}/T_B=2.6$
for nitrogen ($N_2$).
(Other data is given in~\cite{Apfel}.)

In Table below, comparative data for $T_{cr}/4$ are given.

\begin{center}
\begin{tabular}{|c|c|c|c|}
\hline
{Substance} & {$\varepsilon$,  K} & {$T_{cr}/4$} & {$E_{cr}\cdot\varepsilon/k$}\\
\hline
{$Ne$} & 36.3 & 11 & {10.5} \\
\hline
{$Ar$} & 119.3 & 37 & {35}\\
\hline
{$Kr$}& 171 & 52 & {50}\\
\hline
{$N_2$} & 95, 9 & 31 & {28}\\
\hline
{$CH_4$} & 148.2 & 47 & {43}\\
\hline
{$C_2H_6$} & 243.0& 76 & {70} \\
\hline
\end{tabular}

\end{center}

Since, as is well known,
$Z$ begins on the
$(\rho, T)$ coordinates,
we can use the dependence (given above) up to
$Z=0.444$ and then include
the thermic potential,
because we must also take into account
the influence of a third particle.
As was already stated in previous papers,
the dressed, or ``thermic,'' potential~$\Psi(r)$
is attractive.

The ``mixture'' of statistics allowed us to solve the famous problem of Gibbs paradox
as a contradictory example in the system of phenomenological axioms of thermodynamics.
As Poincare noted, we turned to ``arithmetics'' to solve this problem
and used the mixture of the statistic from the number theory and the usual statistic.
In numerous attempts to solve this problem were base on the use of
a mixture of particles~\cite{Gelfer} and their internal structure.
The fact that there are foci at the critical point and the
Zeno line is stable (i.e., the compressibility factor is equal to unity)
allowed us to do this.

This problem was attacked not only by many physicists
but also by philosophers, for the first time, by Poincare in his philosophic works,
and also by B.~V.~Kedrov, S.~D.~Haitun, and many other.

\bigskip
I wish to thank D.~S.~Minenkov and A.~V.~Churkin who performed the computations.
I also wish to express deep gratitude to Professor V.~S.~Vorobiev
for his enthusiasm, for very productive discussions, and for his efforts to verify
the computational results and to compare them with experimental data.
I am cordially grateful to philosopher academician A.~A.~Guseinov for
very useful consultations.

\end{document}